\documentclass[journal]{IEEEtran}
\usepackage{amsmath, bm}
\usepackage{amssymb, amsthm}
\usepackage{authblk, algorithm, algorithmic}
\usepackage{hyperref}
\usepackage{setspace}
\usepackage{url} 
\usepackage{enumitem}
\usepackage{xcolor}
\usepackage{authblk}
\usepackage{bbm}
\usepackage{rotating}
\usepackage{pdflscape}
\usepackage{enumitem}  
\usepackage{subcaption}
\DeclareCaptionLabelSeparator{mysep}{.\quad}
\usepackage{multirow, tabularx}
\usepackage{cite}
\newcommand{\argmax}{\operatornamewithlimits{argmax}}
\newcommand{\bz}{\boldsymbol{z}}
\newcommand{\bw}{\boldsymbol{w}}
\newcommand{\btheta}{\boldsymbol{\theta}}

\ifCLASSINFOpdf
\else
   \usepackage[dvips]{graphicx}
\fi
\usepackage{url}
\usepackage{graphicx}

\newcommand{\numbereqn}{\addtocounter{equation}{1}\tag{\theequation}} 

\begin{document}

\title{Fast Bootstrapping Nonparametric Maximum Likelihood for Latent Mixture Models}

\author{\IEEEauthorblockN{Shijie Wang, Minsuk Shin, and Ray Bai}
\thanks{Manuscript received January 7, 2024; revised February 28, 2024. The work of S. Wang and R. Bai was supported by the U.S. National Science Foundation under Grant DMS-2015528. 

The authors S. Wang and R. Bai are with the Department of Statistics, University of South Carolina, SC 29208 USA (e-mail: \href{mailto:shiejiew@email.sc.edu}{shijiew@email.sc.edu}; \href{mailto:rbai@mailbox.sc.edu}{rbai@mailbox.sc.edu}). M. Shin is with Gauss Labs, CA 94301 USA (e-mail: \href{mailto:shiejiew@email.sc.edu}{minsuk000@gmail.com}).

This paper supersedes a previously circulated technical report by S. Wang and M. Shin (https://arxiv.org/pdf/2006.00767v2.pdf).}}

\maketitle

\begin{abstract}
Estimating the mixing density of a latent mixture model is an important task in signal processing. Nonparametric maximum likelihood estimation is one popular approach to this problem. If the latent variable distribution is assumed to be continuous, then bootstrapping can be used to approximate it. However, traditional bootstrapping requires repeated evaluations on resampled data and is not scalable.  In this letter, we construct a generative process to rapidly produce nonparametric maximum likelihood bootstrap estimates. Our method requires only a single evaluation of a novel two-stage optimization algorithm. Simulations and real data analyses demonstrate that our procedure accurately estimates the mixing density with little computational cost even when there are a hundred thousand observations.
\end{abstract}


\IEEEpeerreviewmaketitle

\section{Introduction}

\IEEEPARstart{N}{onparametric} maximum likelihood estimation (\emph{NPMLE}) is a popular methodology in signal processing applications such as pattern recognition \cite{huynh2020nonparametric, li2021boosting}, channel estimation \cite{bhatia2007non}, signal recovery \cite{feng2017nonparametric}, and positron emission tomography \cite{Silverman1990JRSSB, Vardi1985JASA}. NPMLE also has applications in empirical Bayes and regression modeling \cite{efron2016empirical, jiang2021nonparametric, fan2023EB}.
 Suppose that we observe $\boldsymbol{y} = (y_1, \ldots, y_n)$ where 
\begin{equation} \label{eq:mixture-eb}
    y_i \mid \theta_i \sim f(y_i \mid \theta_i),~~\theta_i \sim \pi(\theta),~~i = 1, \ldots, n,
\end{equation} 
and the density $f$ is known but $\pi$ is unknown. NPMLE aims to estimate the mixing density (or the prior) $\pi$.

The NPMLE estimator $\widehat{\pi}$ in the latent mixture model \eqref{eq:mixture-eb} \cite{fruhwirth2006finite,mclachlan2019finite} was introduced by \cite{kiefer1956consistency} and solves the optimization,
\begin{align*}\label{eq:npmlesol}
\widehat{\pi} & = \underset{\pi\in\mathrm{\Pi}}{\mathrm{argmax}} \sum_{i=1}^n\mathrm{log}[\mathbb{E}_{\pi}\{  f(y_i\mid \theta_i )\} ] \\
& = \underset{\pi\in\mathrm{\Pi}}{\mathrm{argmax}} \sum_{i=1}^n\mathrm{log}  \left\{ \int_{\Theta}  f(y_i\mid \theta_i) d \pi (\theta_i) \right\}, \numbereqn 
\end{align*}
where $\mathrm{\Pi}$ denotes the family of all probability distributions on the parameter space $\Theta$. Under mild regularity conditions, \cite{lindsay1995mixture} showed that the solution $\widehat{\pi}$ of \eqref{eq:npmlesol} exists and is unique, and even if the true $\pi$ is continuous, $\widehat{\pi}$ is almost surely discrete with a support of at most $n$ points. Exploiting this property, there exist many algorithms to solve \eqref{eq:npmlesol}, including the EM algorithm \cite{laird1978nonparametric, zhang2003compound} and convex optimizers \cite{koenker2014convex, feng2018approximate}.

Nevertheless, a discrete NPMLE estimator $\widehat{\pi}$ is unsatisfactory when the latent distribution is reasonably assumed to be continuous \cite{Silverman1990JRSSB, liu2009functional}. 
As a result, smoothed variants of NPMLE have been proposed \cite{Silverman1990JRSSB, efron2016empirical, li2021boosting, koenker2019comment}. However, these methods typically require careful tuning of smoothing parameters such as bandwidth \cite{koenker2019comment, li2021boosting}, roughness penalty term \cite{Silverman1990JRSSB, efron2016empirical}, and/or spline degrees of freedom \cite{li2021boosting, efron2016empirical}. 



As an alternative to smoothing, bootstrapping has been shown to be an effective way to simulate from both prior and posterior continuous densities \cite{Knott2007, rubin1981bayesian, newton1994approximate, Newton2021}. However, bootstrapping is seldomly employed for NPMLE. This may be because the bootstrap traditionally involves resampling the data with replacement and repeatedly optimizing weighted objective functions.
In the context of NPMLE, the standard bootstrap procedure requires one to repetitively optimize 
\begin{equation} \label{eq:bootstrap-NPMLE}
    \widehat{\pi}^{(b)} = \argmax_{\pi} \sum_{i=1}^nw_i^{(b)}\mathrm{log}[\mathbb{E}_{\pi}\{  f(y_i \mid \theta_i )\} ], ~~ \boldsymbol{w}^{(b)}\overset{\text{iid}}\sim \mathbb{P}_{\boldsymbol{w}},
\end{equation}
for some probability measure $\mathbb{P}_{\bw}$. For example, if $\bw \sim \text{Multinomial}(n, \mathbbm{1}_n/n)$, where $\mathbbm{1}_n$ denotes an $n$-dimensional one vector, we have the nonparametric bootstrap \cite{efron1979bootstrap}. If $\bw \sim n \times \text{Dirichlet}(n, \mathbbm{1}_n)$, we have the weighted likelihood bootstrap \cite{newton1994approximate}. Repeatedly solving \eqref{eq:bootstrap-NPMLE} can be time-consuming.

To resolve this computational bottleneck and broaden the applicability of bootstrapping, we build upon the generative bootstrap sampling (GBS) approach of \cite{shin2020generative}. We introduce a new generative framework called Generative Bootstrapping for NPMLE (\emph{GB-NPMLE}). Our contributions are as follows:
\begin{enumerate}
    \item We propose GB-NPMLE for estimating a continuous prior density via bootstrapping. Different from GBS which only requires weights $\bw$ as inputs \cite{shin2020generative}, GB-NPMLE needs \emph{additional} noise inputs $\bz$ to capture the latent representation of $\pi$ in \eqref{eq:mixture-eb}. 
    \item To optimize GB-NPMLE, we introduce a novel two-stage algorithm. In contrast to traditional bootstrapping, GB-NPMLE requires only a \emph{single} evaluation of this two-stage algorithm instead of repetitive evaluations of \eqref{eq:bootstrap-NPMLE}.
\end{enumerate}




\section{Proposed Method}

\subsection{GB-NPMLE}


Recent developments in generative learning such as variational autoencoders (VAEs) \cite{kingma2013auto} and generative adversarial networks (GANs) \cite{goodfellow2014generative,zhou2023deep,liu2021wasserstein} provide another perspective for NPMLE.
Instead of estimating $\pi$, we can construct a generator $T$ that matches to the target density $\pi$, i.e. $T(\bz)\sim \pi$ where $\bz$ $\in \mathbb{R}^q$ follows a known distribution such as a Gaussian or a uniform.
We then reformulate NPMLE optimization \eqref{eq:npmlesol} via a generative framework,
\begin{equation}\label{eq:npmle-gen}
\widehat T= \argmax_{T\in \mathcal{T}}  \sum_{i=1}^n   \mathrm{ log } [\mathbb{E}_{\bz}\{f(y_i\mid T({\bz})\}],
\end{equation}
where ${\bz}\in \mathbb{R}^q$ follows (for example) a standard uniform.
\cite{qiu2021almond} considered \eqref{eq:npmle-gen} where the generator $T$ was constructed by a VAE \cite{kingma2013auto} with a Langevin bias-correction algorithm \cite{roberts1996exponential}.
Unfortuantely, both the classical NPMLE $\widehat{\pi}$ in \eqref{eq:npmlesol} and the generative NPMLE $\widehat{T}$ in \eqref{eq:npmle-gen} result in discrete solutions \cite{lindsay1995mixture}.
This discreteness is less attractive when the true latent $\pi$ is continuous, as in many signal processing applications \cite{Silverman1990JRSSB, Vardi1985JASA}.

To efficiently obtain a bootstrapped NPMLE distribution as a smooth estimator for $\pi$ in \eqref{eq:mixture-eb}, we introduce GB-NPMLE. Our framework (approximately) generates NPMLE bootstrap estimates via a generator function. The generator $G := G(\bw, \bz)$ takes \emph{both} noise $\bz \in$ $\mathbb{R}^q$ \emph{and} bootstrap weights $\bw \in$ $\mathbb{R}^n$ as inputs. We construct $G$ using a feedforward neural network (FNN) \cite{hornik1989multilayer, hornik1991approximation}, because of the FNN's universal approximation properties for a large class of functions \cite{hornik1989multilayer, hornik1991approximation}. 

GB-NPMLE incorporates the GBS approach of \cite{shin2020generative}.
However, GBS is only geared towards uncertainty quantification of a single \emph{point estimate}. Thus, GBS takes \emph{only} weights $\bw$ as inputs and \emph{cannot} be directly used to learn an entire \emph{probability density} $\pi$ as in \eqref{eq:bootstrap-NPMLE}. In order to learn an unknown density $\pi$, GB-NPMLE \emph{also} requires noise inputs $\bz$, as in \eqref{eq:npmle-gen}. 
In summary, GB-NPMLE approximates the bootstrap distribution for NPMLE with the objective function,
\begin{equation}\label{eq:GB-NPMLE}
\widehat G = \argmax_{G} \  \mathbb{E}_{\boldsymbol w}\Big\lbrack \sum_{i=1}^n w_i \ \mathrm{ log } [\mathbb{E}_{\boldsymbol z}\{f(y_i\mid G({\boldsymbol w, \boldsymbol z})\}] \Big\rbrack,
\end{equation}
where $G(\bw,\bz):$ $\mathbb{R}^{n+q} \mapsto \mathbb{R}$. Once we have optimized $G$ (i.e. the weights and biases of the FNN), it is then effortless to generate novel NPMLE bootstrap estimates as $\widehat{G}(\bw_{\text{new}}, \bz_{\text{new}})$.

To optimize \eqref{eq:GB-NPMLE} in practice, the expectations $\mathbb{E}_{\boldsymbol z}$ and $\mathbb{E}_{\boldsymbol w}$ are approximated by Monte Carlo averaging, similarly as in VAEs \cite{kingma2013auto} and GANs \cite{goodfellow2014generative}. Specifically, we independently sample $z \sim \text{Unif}(0,1)$ and $\boldsymbol{w}$ $\sim n\times \text{Dirichlet}(n, \mathbbm{1}_n)$ a sufficiently large number of times and then approximate the expectations in \eqref{eq:GB-NPMLE} with sample averages.

Directly optimizing $G$ in \eqref{eq:GB-NPMLE} using stochastic gradient descent (SGD) \cite{EmmertStreib2020FrontiersinAI} is a nontrivial task. In practice, the Monte Carlo approximation for $\mathbb{E}_{\bz}$ introduces high variance to the approximate gradient for SGD because the log functions in \eqref{eq:GB-NPMLE} greatly shatter the linearity of $\mathbb{E}_{\boldsymbol z}$ \cite{pmlr-v54-lian17a}.
Consequently, this can lead to slow convergence when optimizing the GB-NPMLE objective function \eqref{eq:GB-NPMLE} \cite{pmlr-v54-lian17a}. We address these estimation difficulties with a novel two-stage algorithm. 


\subsection{GB-NPMLE Two-Stage Algorithm}
\label{sec:GB-NPMLE-two}

To efficiently optimize $G$ in \eqref{eq:GB-NPMLE}, we generalize the generator output $\theta = G(\bw, \bz)$ to be a multi-dimensional $\btheta \in \mathbb{R}^{l}$ where $l>1$ is the number of realizations of $\theta$.
The resulting sequence $\btheta = \{ \theta^{(j)} \}_{j=1}^{l}$ serves as the bootstrap sample \emph{candidates}. As we explain shortly, multiple realizations of $\theta$ help to stabilize the gradient approximation in SGD for the Monte Carlo estimate of $\mathbb{E}_{\boldsymbol z}$ in \eqref{eq:GB-NPMLE}.
Unfortunately, this generalization also leads to correlation across the  bootstrap samples.

One way to overcome the correlation issue is to take a random draw of one $\theta^{(j)}$ from $\btheta$ according to mixing probabilities $\boldsymbol{\tau} = (\tau_1, \ldots, \tau_l)$ for the entries of $\btheta$.  Repeating this procedure $B$ times produces $B$ independent bootstrap samples.
However, in addition to generator $G$, we now \emph{also} need to estimate $\boldsymbol{\tau}$. Accordingly, we propose a two-stage algorithm where in Stage I, we train $G$, and in Stage II, we estimate $\boldsymbol{\tau}$.

\vspace{0.5cm}
\noindent{\emph{\textbf{GB-NPMLE Algorithm Stage I:}} Fix $\boldsymbol{\tau} = (\tau_1,...,\tau_l)$ where $\tau_1=...=\tau_l=1/l$, and solve the modified GB-NPMLE objective function,}
\begin{equation}\label{alg:stage I}
\widehat G = \argmax_{G}  \mathbb{E}_{\boldsymbol w}\Big\lbrack \sum_{i=1}^n w_i \ \mathrm{ log } [\mathbb{E}_{\boldsymbol z, \gamma}\{f(y_i\mid \boldsymbol{e}_{\gamma}^{\top} G({\boldsymbol w, \boldsymbol z})\}] \Big\rbrack,
\end{equation}
where $\gamma \sim$ Multinomial$(1,\boldsymbol{\tau})$, $\boldsymbol{e}_{\gamma}$ is the $\gamma$-th unit vector, and $\boldsymbol{e}_{\gamma}^{\top} G({\boldsymbol w, \boldsymbol z}) \in \mathbb{R}$ is the $\gamma$-th entry of generator output $\btheta$.


To see how the modified GB-NPMLE objective \eqref{alg:stage I} helps to relieve high Monte Carlo variance, let $\mathbb{E}_{\boldsymbol z, \gamma}\{f(y_i | \boldsymbol{e}_{\gamma}^{\top} G({\boldsymbol w, \boldsymbol z})\}$ in \eqref{alg:stage I} be denoted by $\mathbb{E}_{\boldsymbol z} \{ \mathbb{E}_{\gamma} \{ f(\btheta[\bw,\bz,\gamma]) \} \}$, and similarly, let $\mathbb{E}_{\boldsymbol z} \{ f(y_i|G(\bw,\bz)) \}$ be $\mathbb{E}_{\boldsymbol z} \{ f(\btheta[\bw,\bz]) \}$ in \eqref{eq:GB-NPMLE}. By the law of total variance, $\text{Var}\{ f(\btheta[\bw,\bz]) \} \geq \text{Var} \{ \mathbb{E}_{\gamma}f(\btheta[\bw,\bz,\gamma]) \}$. Hence, the generalization of the generator output dimension to $l > 1$ reduces Monte Carlo variance.

\vspace{0.4cm}
\noindent \emph{\textbf{GB-NPMLE Algorithm Stage II:}} Fix the well-trained generator $\widehat{G}$ from \eqref{alg:stage I} and estimate $\boldsymbol{\tau}$ via the Monte Carlo EM (MCEM) algorithm \cite{levine2001implementations}. In particular, the MCEM updates for each  $k$th entry of $\boldsymbol{\tau}$ have the closed form,
\begin{equation}\label{alg:stage II}
\tau_{k}^{(t+1)} = \frac{1}{n} \sum_{i=1}^n \frac{\tau_{k}^{(t)} \cdot \mathbb{E}_{\bw,\bz} \{ f(y_i\mid \boldsymbol{e}_{k}^\top \widehat{G}({\bw},\bz) \} }{\sum_{k=1}^{l}\tau_{k}^{(t)} \cdot  \mathbb{E}_{\bw,\bz} \{ f(y_i\mid \boldsymbol{e}_{k}^\top \widehat{G}({\bw},\bz) \} }.
\end{equation}
Note that we \emph{already} obtained an efficient estimator of $\btheta$ from Stage I, and hence, the MCEM algorithm tends to have very fast convergence. In the literature, slow convergence of MCEM mostly lies in obtaining a sequence of estimators $\btheta$ \cite{qiu2021almond} . We circumvent this issue by \emph{fixing} the generator from Stage I.

Once we have estimated both $G$ (in Stage I) and $\boldsymbol{\tau}$ (in Stage II), we can easily generate a new bootstrap estimate by sampling $\boldsymbol{w}_{\text{new}}$, $\bz_{\text{new}}$, $\gamma_{\text{new}}$, and then taking $\theta_{\text{new}} = \boldsymbol{e}_{\gamma_{\text{new}}}^\top \widehat{G}(\boldsymbol{w}_{\text{new}}, \bz_{\text{new}})$. Repeating this procedure $B$ times results in $B$ independent bootstrap estimates. Thus, in contrast to bootstrapped NPMLE which requires $B$ total repetitive evaluations of \eqref{eq:bootstrap-NPMLE}, GB-NPMLE requires only a \textit{single} evaluation of the two-stage algorithm. The entire two-stage GB-NPMLE procedure is given in Algorithm \ref{alg:gb-npmle}. Empirical evidence (reported in Appendix \ref{sec:cov}) demonstrates sufficient convergence of our algorithm.


\begin{figure*}
\centering
\includegraphics[width=.93\textwidth]{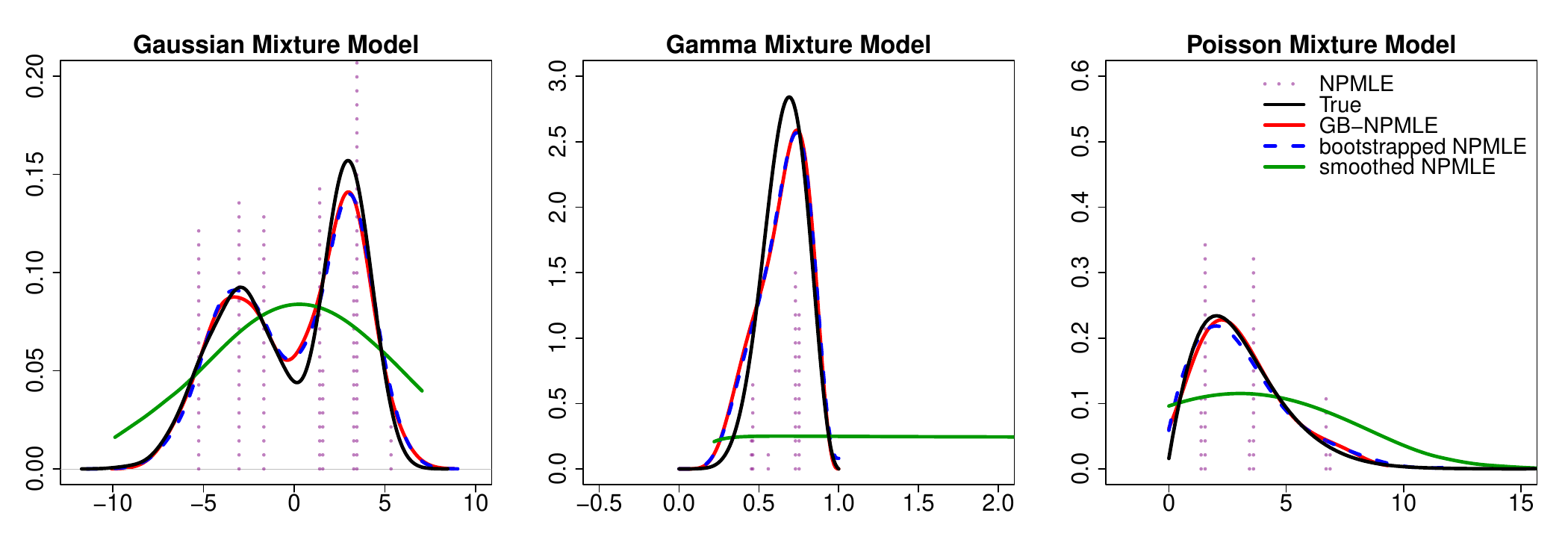}
\caption{Results from one replication of Simulations (i)-(iii). In addition to GB-NPMLE (solid red), bootstrapped NPMLE (dashed blue), and smoothed NPMLE (solid green), we also plot the true density (solid black) and the classical discrete NPMLE (dotted purple).}\label{fig:sim}
\end{figure*}

\begin{algorithm}[tbh] \
\footnotesize
  \caption{\footnotesize GB-NPMLE Two-stage Algorithm }\label{alg:gb-npmle}
  \begin{algorithmic}[1]
  \STATE \emph{Set} $l=100$, $tol=0.001$, $T = 2000$, $B=1000$.
  \STATE \emph{Initialize} neural network parameters $\phi$ and $\boldsymbol{\tau} = (1/l, \ldots, 1/l)^\top$
      \STATE \textbf{procedure:} Stage I
      \begin{ALC@g}
      \FOR{$t$ in $1,\ldots,\text{Epoch } T$}
        \STATE \textit{Sample} $\boldsymbol{w}\sim n\times \text{Dirichlet}(n, \mathbbm{1}_n)$
        \STATE \textit{Sample} $z \sim \text{Unif}$(0,1)
        \STATE \textit{Sample} $\gamma \sim$ Multinomial$(1,\boldsymbol{\tau})$
        \STATE \textit{Update} $G_{\phi}$ by SGD using modified GB-NPMLE objective \eqref{alg:stage I}
        
        
      \ENDFOR
      \STATE \textbf{return} $G_{\phi}$
      \end{ALC@g}
    \STATE \textbf{end procedure}
    
    \STATE \textbf{procedure:} Stage II
    \begin{ALC@g}
      \STATE \emph{Initialize} $\boldsymbol{\tau}^{new} = ( 1/l, \ldots, 1/l )$ and $\boldsymbol{\tau}^{old} = (0,\ldots,0)$
      \WHILE{$\min_k(|\tau_k^{old} - \tau_k^{new}|) \geq tol$}
        \STATE \textit{set} $\boldsymbol{\tau}^{old}$ = $\boldsymbol{\tau}^{new}$

        \FOR{$k$ in $1,\ldots, l$}
        \STATE \textit{Sample} $\boldsymbol{w}\sim n\times \text{Dirichlet}(n, \mathbbm{1}_n)$
        \STATE \textit{Sample} $z \sim \text{Unif}$(0,1)
        
        \STATE \textit{Update} $\tau_k^{new}$ as in \eqref{alg:stage II}
                
        \ENDFOR
        
      \ENDWHILE
      \STATE \textbf{return} $\boldsymbol{\tau}^{new}$
      \end{ALC@g}
    \STATE \textbf{end procedure}

    \STATE \textbf{procedure:} Generate $B$ bootstrap estimates
    \begin{ALC@g}
    \FOR{$b$ in $1, \ldots, B$}
    \STATE \textit{Sample} $\boldsymbol{w}^{(b)}\sim n\times \text{Dirichlet}(n, \mathbbm{1}_n)$
        \STATE \textit{Sample} $z^{(b)} \sim \text{Unif}$(0,1)
        \STATE \textit{Sample} $\gamma^{(b)} \sim$ Multinomial$(1, \boldsymbol{\tau}^{new})$
        \STATE \textit{Generate} $\theta^{(b)} = e_{\gamma^{(b)}}^\top  \widehat{G}(\boldsymbol{w}^{(b)}, z^{(b)})$
    \ENDFOR
    \STATE \textbf{return} $\theta^{(1)}, \ldots, \theta^{(B)}$

    \end{ALC@g}
    \STATE \textbf{end procedure}
  \end{algorithmic}
\end{algorithm}


\section{Experiments and Results}\label{sec:sim}

\subsection{Simulation Studies}  \label{sec:experiments}
We investigate the performance of GB-NPMLE optimized by the two-stage algorithm from Section \ref{sec:GB-NPMLE-two}. 
The generator $G$ is an FNN with two hidden layers and 500 neurons per layer. Sensitivity analysis to this choice of architecture is presented in Appendix \ref{sec:sensitivity}.
The expectations $\mathbb{E}_{\bw}$, $\mathbb{E}_{\bz}$ and $\mathbb{E}_{\gamma}$ in the modified GB-NPMLE objective \eqref{alg:stage I} are approximated by Monte Carlo averages of $100$ independent samples of $\boldsymbol{w}\sim n\times \text{Dirichlet}(n, \mathbbm{1}_n)$, $z \sim \text{Unif}$(0,1) and $\gamma \sim$ Multinomial$(1,\boldsymbol{\tau})$, and the length of $\boldsymbol{\tau}$ is set to be $l=100$.
To train $G$, we use SGD with the Adam optimizer \cite{kingma2014adam}. All experiments are conducted on a single NVIDIA GeForce RTX 2080 Ti graphics processing unit (GPU) with 11GB RAM.

We consider three simulation settings for the prior $\pi$ in the mixture model \eqref{alg:gb-npmle}: (i) bimodal, (ii) unimodal and bounded on $(0,1)$, and (iii) skewed right. Additional simulations are conducted in Appendix \ref{sec:moresims}.
Here, we consider:
\begin{enumerate}[label=(\roman*)]
    \item  Gaussian mixture model (\textbf{GMM}): $y \mid \theta \sim \mathcal{N}(\theta, 1)$ and $\theta = 0.5\mathcal{N}(-3,2)+0.5\mathcal{N}(3,1)$;
    \item Gamma mixture model (\textbf{GaMM}): $y \mid \theta \sim \text{Gamma}(10,\theta)$ and $\theta \sim \text{Beta}(10, 5)$;
    \item Poisson mixture model (\textbf{PMM}): $y \mid \theta \sim \text{Poisson}(\theta)$ and $\theta \sim \text{Gamma}(3,1)$.
\end{enumerate}

We compare GB-NPMLE to bootstrapped NPMLE \eqref{eq:bootstrap-NPMLE} and a smoothed version of the discrete NPMLE \eqref{eq:npmlesol}. We use $B=10{,}000$ bootstrap estimates to estimate the prior for GB-NPMLE and bootstrapped NPMLE. For bootstrapped NPMLE, we use the nonparametric bootstrap, i.e. we sample $n$ instances of the data with replacement and then optimize \eqref{eq:npmlesol} a total of $B$ times using the \textsf{R} package \texttt{REBayes} \cite{KoenkerGu2017}. For the smoothed NPMLE, we use kernel smoothing of $\widehat{\pi}$ in \eqref{eq:npmlesol} with the \texttt{KWsmooth} function in \texttt{REBayes} \cite{KoenkerGu2017}, where we select the optimal bandwidth from an equispaced grid of length $25$ from $0.1$ to $10$ using leave-one-out cross-validation (LOOCV). 

For performance comparison of the three NPMLE methods, we generate artificial datasets of size $n = 1000$.  We look at: a) Wasserstein-1 distance between $\pi$ and $\widehat{\pi}$, defined as $W_1(\pi, \widehat{\pi})=\int| \mathbb{F}_{\pi}(x)-\mathbb{F}_{\widehat{\pi}}(x)|dx,$ where $\mathbb{F}$ is the cumulative density function (CDF) of $\pi$, and b) Integrated Squared Error (ISE), defined as $\int (\widehat{\pi}(x)-\pi(x))^2dx$.
Lower $W_1(\pi, \widehat{\pi})$ and ISE indicate better performance. Our results for simulations (i)-(iii) averaged across 20 replications are shown in Table \ref{tab:sim}.

\begin{table}[tbh]
\centering
\caption{COMPARISONS OF PERFORMANCE OF DIFFERENT NPMLE METHODS ON SIMULATED DATASETS }\label{tab:sim}
\resizebox{1.0\columnwidth}{!}{
\begin{tabular}{c|cc|cc|cc}
\hline
&\multicolumn{2}{c|}{GB-NPMLE}&
\multicolumn{2}{c|}{Bootstrapped NPMLE}&
\multicolumn{2}{c}{Smoothed NPMLE}\\
\hline
Model&$W_1(\pi,\widehat{\pi})$&ISE&$W_1(\pi,\widehat{\pi})$&ISE&$W_1(\pi,\widehat{\pi})$&ISE\\
\hline
\hline
GMM&0.348&0.008&0.298&0.008&0.936&0.034\\
GaMM&0.032&0.263&0.037&0.497&0.223&1.731\\
PMM&0.400&0.045&0.389&0.036&1.763&0.059\\
\hline
\hline
\end{tabular}}
\end{table}


From Table \ref{tab:sim}, we first observe that GB-NPMLE has a very similar performance to bootstrapped NPMLE. On the other hand, smoothed NPMLE (with bandwidth parameter chosen by LOOCV) performs much worse than the bootstrapping approaches. This demonstrates the difficulty of optimally tuning smoothing parameters for smoothed NPMLE. 

The results from one replication of simulations (i)-(iii) are plotted in Fig. \ref{fig:sim}, which also plots the true latent density $\pi$ (solid black) and the discrete NPMLE solution \eqref{eq:npmlesol} (dotted purple). Fig. \ref{fig:sim} shows that the classical (discrete) NPMLE \eqref{eq:npmlesol} is inadequate for capturing $\pi$ when $\pi$ is truly continuous. For example, in the GaMM model, the classical NPMLE does not clearly indicate that $\pi$ is unimodal. On the other hand, GB-NPMLE (solid red) and bootstrapped NPMLE (solid blue) are quite close to the true $\pi$, capturing all inherent aspects of the true $\pi$ like bimodality, boundedness and unimodality, and skewness respectively in the GMM, GaMM, and PMM models. The smoothed NPMLE appears to be oversmoothed and is unable to capture bimodality of $\pi$ in the GMM model or the unimodality and boundedness of $\pi$ in the GaMM model. In the PMM model, the sharp peak of the skewed density $\pi$ is also not adequately captured by smoothed NPMLE. 



\begin{figure}[t]
\centering
\includegraphics[width=\linewidth]{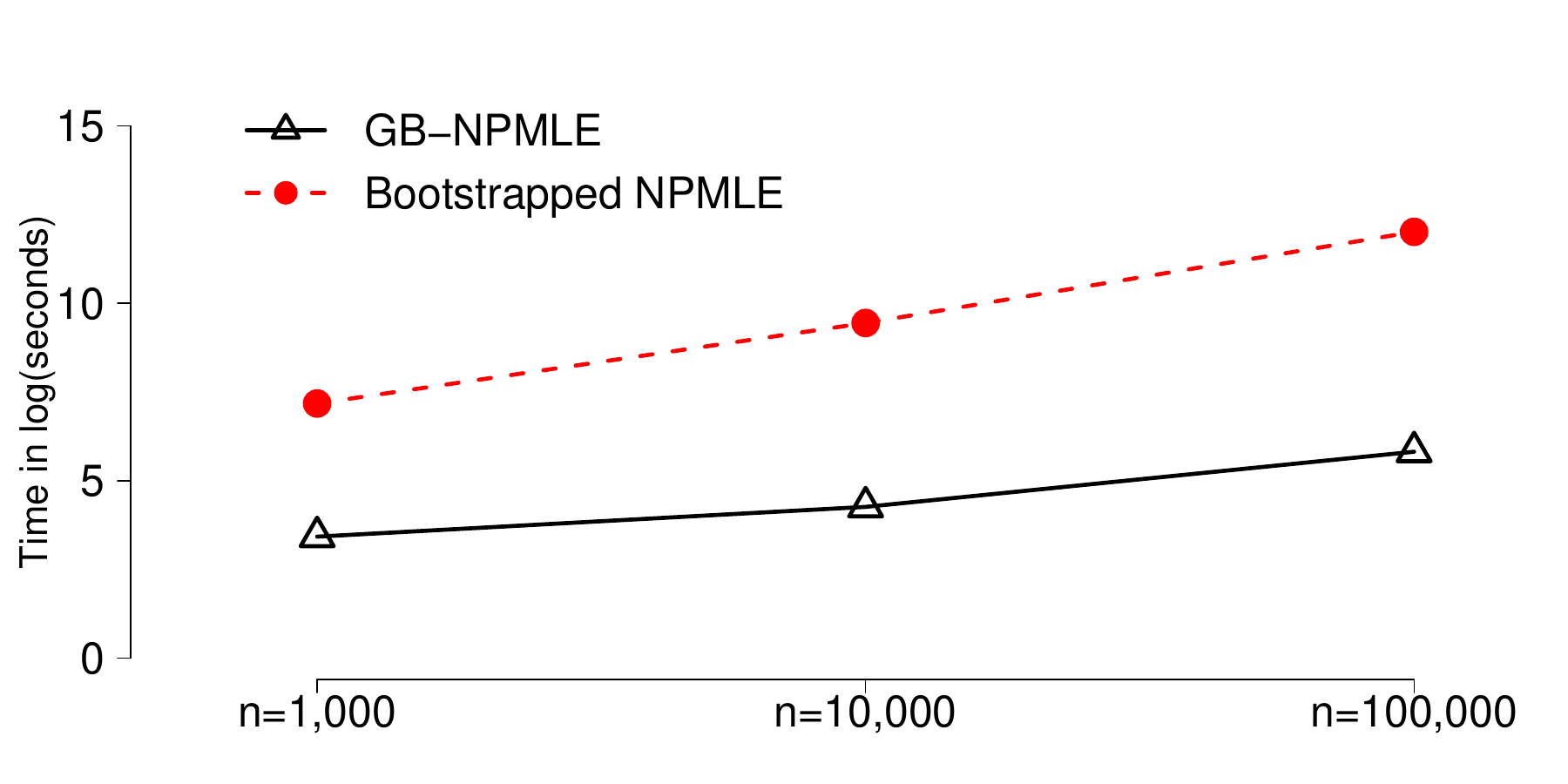}
\caption{{Mean computation time in log(seconds) for the three simulations vs. sample size $n \in \{1000, 10{,}000, 100{,}000 \}$. For $n = 100{,}000$, GB-NPMLE takes five minutes, whereas bootstrapped NPMLE takes almost two days. }}\label{fig:time}
\end{figure}


GB-NPMLE and bootstrapped NPMLE both give similar performance, but GB-NPMLE is much faster because it only requires a single evaluation of Algorithm \ref{alg:gb-npmle}. We compare the average computational time of GB-NPMLE and bootstrapped NPMLE for sample sizes $n \in \{ 1000, 10{,}000, 100{,}000 \}$ across simulations (i)-(iii). Our results are plotted (on the log scale) in Fig. \ref{fig:time}. Fig. \ref{fig:time} shows that GB-NPMLE is much more scalable. In particular, GB-NPMLE takes only about five minutes to complete when $n=100{,}000$; on the other hand, bootstrapped NPMLE requires almost two days to finish. 



\subsection{Evaluation on Real Datasets} \label{sec:real}
We evaluate GB-NPMLE on three count datasets where the counts are assumed to follow a Poisson mixture model, $y_i \mid \theta_i \sim \text{Poisson}(\theta_i)$ and $\theta_i \sim \pi, i = 1, \ldots, n$.
The \emph{Norberg} dataset consists of $1125$ group life insurance statistics in Norway \cite{norberg1989experience}, the \emph{Thailland} dataset consists of 602 preschool children's health condition, and the \emph{Mortality} dataset contains the number of deaths for women aged greater than eighty \cite{bohning1999computer}.
\begin{figure}[t]
\centering
\includegraphics[width=\linewidth]{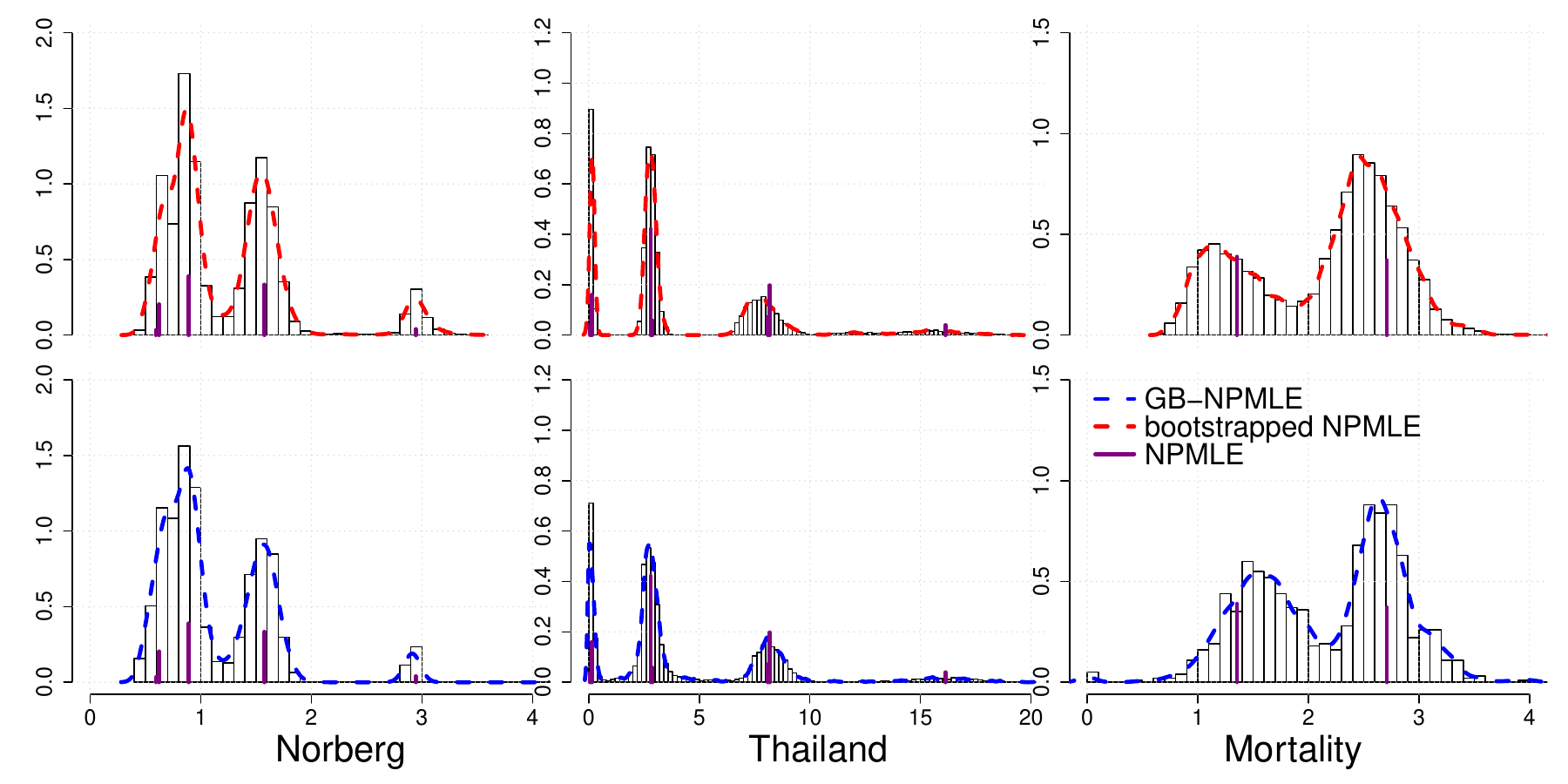}
\caption{Results for real datasets for GB-NPMLE (dashed red, top panel) and bootstrapped NPMLE (dotted blue, bottom panel). The pink vertical lines in both plots represent the discrete NPMLE solution.}\label{fig:real}
\end{figure}
Fig. \ref{fig:real} compares GB-NPMLE (top panel) with bootstrapped NPMLE (bottom panel). We see that GB-NPMLE approximates bootstrapped NPMLE very well. Furthermore, both methods identify the same number of local modes (three in \emph{Norberg}, four in \emph{Thailand}, and two in \emph{Mortality}). 

We also examine out-of-sample prediction using the log predictive score (LPS) $\mathbb{E}_{\theta} [-\log f(\boldsymbol{y}_\text{new};\theta) \mid \boldsymbol{y}_\text{obs}]$. To approximate LPS, we use $K$-fold cross validation (CV),
\begin{equation*}
    \text{LPS} = K^{-1} \sum_{k=1}^{K} \sum_{i\in I_k} -\log \bigg[ B^{-1}\sum_{b=1}^B f(y_i;\widehat{\theta}^{b}_{(-k)}) \bigg],
\end{equation*} 
where $\widehat{\theta}^{b}_{(-k)}\sim\widehat{\pi}_{(-k)}$ and $\widehat{\pi}_{(-k)}$ is the bootstrap distribution obtained by excluding the $k$th fold $I_k$. We set $K=10$ and $B = 500$. Table \ref{tab:real} shows that the GB-NPMLE predictions are quite close to their standard bootstrap counterparts.. Table \ref{tab:real} also shows that GB-NPMLE is much more scalable, with only a modest increase in computation time for 10-fold CV when $n$ increases from 72 to 1096. When $n = 1096$, GB-NPMLE is about 34 times faster than bootstrapped NPMLE.

\begin{table}[tbh]
\caption{\footnotesize COMPARISONS OF PERFORMANCE OF GB-NPMLE AND BOOTSTRAPPED NPMLE ON REAL COUNT DATASETS}\label{tab:real}
\centering
\resizebox{1.0\columnwidth}{!}{%
\begin{tabular}{cc|cc|cc}
\hline
&&\multicolumn{2}{c|}{GB-NPMLE}&\multicolumn{2}{c}{Bootstrapped NPMLE}\\
\hline
Dataset&$n$&LPS&Time (sec)&LPS&Time (sec)\\
\hline
\hline
Norberg&72 & 20.35 & 126.1 & 20.83 &732.8 \\
Thailland & 602 & 156.26 & 140.9 & 156.21&3439.3 \\
Mortality & 1096 & 199.30 & 153.3 & 199.31&5279.2 \\
\hline
\end{tabular}}
\end{table}


\section{Conclusion}
In this letter, we introduced GB-NPMLE, a generative framework for estimating a mixing density through bootstrapping. A novel two-stage algorithm was proposed to rapidly and accurately obtain bootstrap estimates. Bootstrapping has rarely been employed for NPMLE in the past, and our work paves the way for bootstrapping to serve as an attractive alternative to smoothing NPMLE. Simulations and real data analyses
demonstrated the effectiveness and scalability of our approach. Codes to implement GB-NPMLE and the real datasets analyzed in this article are available on GitHub at
\texttt{\url{https://github.com/shijiew97/GBnpmle}}.

Our methodology is only presented for estimating a univariate latent density. However, it will be useful to extend GB-NPMLE to \emph{multivariate} density estimation. If covariates are also observed, then GB-NPMLE can be extended to mixture of regression models \cite{jiang2021nonparametric}. Finally, theoretical analysis of the two-stage algorithm (e.g. the convergence rate) is needed.

\newpage
\bibliographystyle{IEEEtran}
\bibliography{shijie}

\begin{thebibliography}{10}
\providecommand{\url}[1]{#1}
\csname url@samestyle\endcsname
\providecommand{\newblock}{\relax}
\providecommand{\bibinfo}[2]{#2}
\providecommand{\BIBentrySTDinterwordspacing}{\spaceskip=0pt\relax}
\providecommand{\BIBentryALTinterwordstretchfactor}{4}
\providecommand{\BIBentryALTinterwordspacing}{\spaceskip=\fontdimen2\font plus
\BIBentryALTinterwordstretchfactor\fontdimen3\font minus
  \fontdimen4\font\relax}
\providecommand{\BIBforeignlanguage}[2]{{%
\expandafter\ifx\csname l@#1\endcsname\relax
\typeout{** WARNING: IEEEtran.bst: No hyphenation pattern has been}%
\typeout{** loaded for the language `#1'. Using the pattern for}%
\typeout{** the default language instead.}%
\else
\language=\csname l@#1\endcsname
\fi
#2}}
\providecommand{\BIBdecl}{\relax}
\BIBdecl

\bibitem{huynh2020nonparametric}
H.~T. Huynh and L.~Nguyen, ``Nonparametric maximum likelihood estimation using
  neural networks,'' \emph{Pattern Recognition Letters}, vol. 138, pp.
  580--586, 2020.

\bibitem{li2021boosting}
Y.~Li and Z.~Ye, ``Boosting in univariate nonparametric maximum likelihood
  estimation,'' \emph{IEEE Signal Processing Letters}, vol.~28, pp. 623--627,
  2021.

\bibitem{bhatia2007non}
V.~Bhatia and B.~Mulgrew, ``Non-parametric likelihood based channel estimator
  for {G}aussian mixture noise,'' \emph{Signal Processing}, vol.~87, no.~11,
  pp. 2569--2586, 2007.

\bibitem{feng2017nonparametric}
L.~Feng, R.~Ma, and L.~H. Dicker, ``Nonparametric maximum likelihood
  approximate message passing,'' in \emph{2017 51st Annual Conference on
  Information Sciences and Systems (CISS)}.\hskip 1em plus 0.5em minus
  0.4em\relax IEEE, 2017, pp. 1--6.

\bibitem{Silverman1990JRSSB}
B.~W. Silverman, M.~C. Jones, J.~D. Wilson, and D.~W. Nychka, ``A smoothed {EM}
  approach to indirect estimation problems, with particular, reference to
  stereology and emission tomography,'' \emph{Journal of the Royal Statistical
  Society. Series B (Methodological)}, vol.~52, no.~2, pp. 271--324, 1990.

\bibitem{Vardi1985JASA}
Y.~Vardi, L.~A. Shepp, and L.~Kaufman, ``A statistical model for positron
  emission tomography,'' \emph{Journal of the American Statistical
  Association}, vol.~80, no. 389, pp. 8--20, 1985.

\bibitem{efron2016empirical}
B.~Efron, ``Empirical {B}ayes deconvolution estimates,'' \emph{Biometrika},
  vol. 103, no.~1, pp. 1--20, 2016.

\bibitem{jiang2021nonparametric}
H.~Jiang and A.~Guntuboyina, ``A nonparametric maximum likelihood approach to
  mixture of regression,'' \emph{arXiv preprint arXiv:2108.09816}, 2021.

\bibitem{fan2023EB}
Z.~Fan, L.~Guan, Y.~Shen, and Y.~Wu, ``Gradient flows for empirical {B}ayes in
  high-dimensional linear models,'' \emph{arXiv preprint arXiv:2312.12708},
  2023.

\bibitem{fruhwirth2006finite}
S.~Fr{\"u}hwirth-Schnatter, \emph{Finite Mixture and Markov Switching
  Models}.\hskip 1em plus 0.5em minus 0.4em\relax Springer, 2006.

\bibitem{mclachlan2019finite}
G.~J. McLachlan, S.~X. Lee, and S.~I. Rathnayake, ``Finite mixture models,''
  \emph{Annual Review of Statistics and Its Application}, vol.~6, pp. 355--378,
  2019.

\bibitem{kiefer1956consistency}
J.~Kiefer and J.~Wolfowitz, ``Consistency of the maximum likelihood estimator
  in the presence of infinitely many incidental parameters,'' \emph{The Annals
  of Mathematical Statistics}, vol.~27, no.~4, pp. 887--906, 1956.

\bibitem{lindsay1995mixture}
B.~G. Lindsay, ``Mixture models: Theory, geometry and applications,'' in
  \emph{NSF-CBMS Regional Conference Series in Probability and Statistics},
  vol.~5, 1995, pp. 1--163.

\bibitem{laird1978nonparametric}
N.~Laird, ``Nonparametric maximum likelihood estimation of a mixing
  distribution,'' \emph{Journal of the American Statistical Association},
  vol.~73, no. 364, pp. 805--811, 1978.

\bibitem{zhang2003compound}
C.~H. Zhang, ``Compound decision theory and empirical {B}ayes methods,''
  \emph{Annals of Statistics}, vol.~31, no.~2, pp. 379--390, 2003.

\bibitem{koenker2014convex}
R.~Koenker and I.~Mizera, ``Convex optimization, shape constraints, compound
  decisions, and empirical {B}ayes rules,'' \emph{Journal of the American
  Statistical Association}, vol. 109, no. 506, pp. 674--685, 2014.

\bibitem{feng2018approximate}
L.~Feng and L.~H. Dicker, ``Approximate nonparametric maximum likelihood for
  mixture models: A convex optimization approach to fitting arbitrary
  multivariate mixing distributions,'' \emph{Computational Statistics \& Data
  Analysis}, vol. 122, pp. 80--91, 2018.

\bibitem{liu2009functional}
L.~Liu, M.~Levine, and Y.~Zhu, ``A functional {EM} algorithm for mixing density
  estimation via nonparametric penalized likelihood maximization,''
  \emph{Journal of Computational and Graphical Statistics}, vol.~18, no.~2, pp.
  481--504, 2009.

\bibitem{koenker2019comment}
R.~Koenker and J.~Gu, ``Comment: Minimalist g-modeling,'' \emph{Statistical
  Science}, vol.~34, no.~2, pp. 209--213, 2019.

\bibitem{Knott2007}
M.~Knott and P.~Tzamourani, ``Bootstrapping the estimated latent distribution
  of the two-parameter latent trait model,'' \emph{British Journal of
  Mathematical and Statistical Psychology}, vol.~60, no.~1, pp. 175--191, 2007.

\bibitem{rubin1981bayesian}
D.~B. Rubin, ``The {B}ayesian bootstrap,'' \emph{The Annals of Statistics},
  vol.~9, no.~1, p. 130434, 1981.

\bibitem{newton1994approximate}
M.~A. Newton and A.~E. Raftery, ``Approximate {B}ayesian inference with the
  weighted likelihood bootstrap,'' \emph{Journal of the Royal Statistical
  Society: Series B (Methodological)}, vol.~56, no.~1, pp. 3--26, 1994.

\bibitem{Newton2021}
M.~A. Newton, N.~G. Polson, and J.~Xu, ``Weighted {B}ayesian bootstrap for
  scalable posterior distributions,'' \emph{Canadian Journal of Statistics},
  vol.~49, no.~2, pp. 421--437, 2021.

\bibitem{efron1979bootstrap}
B.~Efron, ``Bootstrap methods: Another look at the jackknife,'' \emph{The
  Annals of Statistics}, vol.~7, no.~1, pp. 1--26, 1979.

\bibitem{shin2020generative}
M.~Shin, S.~Wang, and J.~S. Liu, ``Generative multiple-purpose sampler for
  weighted {M}-estimation,'' \emph{Journal of Computational and Graphical
  Statistics (to appear)}, 2023.

\bibitem{kingma2013auto}
D.~P. Kingma and M.~Welling, ``Auto-encoding variational {b}ayes,'' \emph{arXiv
  preprint arXiv:1312.6114}, 2013.

\bibitem{goodfellow2014generative}
I.~Goodfellow, J.~Pouget-Abadie, M.~Mirza, B.~Xu, D.~Warde-Farley, S.~Ozair,
  A.~Courville, and Y.~Bengio, ``Generative adversarial nets,'' \emph{Advances
  in Neural Information Processing Systems}, vol.~27, 2014.

\bibitem{zhou2023deep}
X.~Zhou, Y.~Jiao, J.~Liu, and J.~Huang, ``A deep generative approach to
  conditional sampling,'' \emph{Journal of the American Statistical
  Association}, vol. 118, no. 543, pp. 1837--1848, 2023.

\bibitem{liu2021wasserstein}
S.~Liu, X.~Zhou, Y.~Jiao, and J.~Huang, ``Wasserstein generative learning of
  conditional distribution,'' \emph{arXiv preprint arXiv:2112.10039}, 2021.

\bibitem{qiu2021almond}
Y.~Qiu and X.~Wang, ``{ALMOND}: Adaptive latent modeling and optimization via
  neural networks and langevin diffusion,'' \emph{Journal of the American
  Statistical Association}, vol. 116, no. 535, pp. 1224--1236, 2021.

\bibitem{roberts1996exponential}
G.~O. Roberts and R.~L. Tweedie, ``Exponential convergence of {L}angevin
  distributions and their discrete approximations,'' \emph{Bernoulli}, vol.~2,
  no.~4, pp. 341--363, 1996.

\bibitem{hornik1989multilayer}
K.~Hornik, M.~Stinchcombe, and H.~White, ``Multilayer feedforward networks are
  universal approximators,'' \emph{Neural Networks}, vol.~2, no.~5, pp.
  359--366, 1989.

\bibitem{hornik1991approximation}
K.~Hornik, ``Approximation capabilities of multilayer feedforward networks,''
  \emph{Neural networks}, vol.~4, no.~2, pp. 251--257, 1991.

\bibitem{EmmertStreib2020FrontiersinAI}
F.~Emmert-Streib, Z.~Yang, H.~Feng, S.~Tripathi, and M.~Dehmer, ``An
  introductory review of deep learning for prediction models with big data,''
  \emph{Frontiers in Artificial Intelligence}, vol.~3, 2020.

\bibitem{pmlr-v54-lian17a}
X.~Lian, M.~Wang, and J.~Liu, ``Finite-sum composition optimization via
  variance reduced gradient descent,'' in \emph{Proceedings of the 20th
  International Conference on Artificial Intelligence and Statistics}, A.~Singh
  and J.~Zhu, Eds., vol.~54.\hskip 1em plus 0.5em minus 0.4em\relax PMLR,
  20--22 Apr 2017, pp. 1159--1167.

\bibitem{levine2001implementations}
R.~A. Levine and G.~Casella, ``Implementations of the {M}onte {C}arlo {EM}
  algorithm,'' \emph{Journal of Computational and Graphical Statistics},
  vol.~10, no.~3, pp. 422--439, 2001.

\bibitem{kingma2014adam}
D.~P. Kingma and J.~Ba, ``Adam: A method for stochastic optimization,''
  \emph{arXiv preprint arXiv:1412.6980}, 2014.

\bibitem{KoenkerGu2017}
R.~Koenker and J.~Gu, ``{REBayes}: An {R} package for empirical {B}ayes mixture
  methods,'' \emph{Journal of Statistical Software}, vol.~82, no.~8, p. 1–26,
  2017.

\bibitem{norberg1989experience}
R.~Norberg, ``Experience rating in group life insurance,'' \emph{Scandinavian
  Actuarial Journal}, vol. 1989, no.~4, pp. 194--224, 1989.

\bibitem{bohning1999computer}
D.~B{\"o}hning, \emph{Computer-Assisted Analysis of Mixtures and Applications:
  Meta-analysis, disease mapping and others}.\hskip 1em plus 0.5em minus
  0.4em\relax CRC Press, 1999.

\end{thebibliography}

\newpage

\appendix

\subsection{Sensitivity Analysis for Generator Architecture} \label{sec:sensitivity}
We conduct a sensitivity analysis to study whether different choices for the number of hidden layers $L$ or the number of hidden neurons $h$ in the generator $G$ affects GB-NPMLE's performance. We first fix $L = 2$ and vary $h \in \{ 50, 250, 500, 750\}$. Next, we fix $h=500$ and vary $L \in \{1, 2, 3, 4 \}$. 

We repeat all of the simulations described in Section \ref{sec:experiments} 20 times and record the average performance metrics (i.e. $W_1(\pi, \widehat{\pi})$ and ISE). Our results are summarized in Table \ref{tab:NN-node-sim}. We do not observe any significant differences in the performance of GB-NPMLE for these different combinations of $(L, h)$. In practice, we recommend a default choice of $L=2$ and $h=500$ for satisfactory performance.
\begin{table}[h]
	\centering
	\caption{\small SENSITIVITY ANALYSIS RESULTS}
	\label{tab:NN-node-sim}
	\resizebox{1.0\columnwidth}{!}{%
		\begin{tabular}{c|cc|cc|cc|cc|}
			\hline
			$L=2$\phantom{1!} &\multicolumn{2}{c|}{$h=50$}&\multicolumn{2}{c|}{$h=250$}&\multicolumn{2}{c|}{$h=500$}&\multicolumn{2}{c|}{$h=750$}\\
			\hline
			Model&$W_1(\pi,\widehat{\pi})$&$\text{ISE}$&$W_1(\pi,\widehat{\pi})$&$\text{ISE}$&$W_1(\pi,\widehat{\pi})$&$\text{ISE}$&$W_1(\pi,\widehat{\pi})$&$\text{ISE}$
			\\
			\hline
			\hline
			GMM&0.334&0.008&0.326&0.008&0.348&0.008&0.329&0.008\\
			GaMM&0.032&0.223&0.032&0.282&0.032&0.263&0.032&0.245\\
			PMM&0.413&0.033&0.380&0.028&0.400&0.045&0.386&0.030\\
			\hline
	\end{tabular}}
	\vspace{.2cm} 
	
	\resizebox{1.0\columnwidth}{!}{%
		\begin{tabular}{c|cc|cc|cc|cc|}
			\hline
			$h=500$ &\multicolumn{2}{c|}{$L=1$}&\multicolumn{2}{c|}{$L=2$}&\multicolumn{2}{c|}{$L=3$}&\multicolumn{2}{c|}{$L=4$}\\
			\hline
			Model&$W_1(\pi,\widehat{\pi})$&$\text{ISE}$&$W_1(\pi,\widehat{\pi})$&$\text{ISE}$&$W_1(\pi,\widehat{\pi})$&$\text{ISE}$&$W_1(\pi,\widehat{\pi})$&$\text{ISE}$
			\\
			\hline
			\hline
			GMM&0.330&0.008&0.348&0.008&0.331&0.008&0.332 &0.008\\
			GaMM&0.032&0.292&0.032&0.263&0.032 &0.304 &0.032 &0.270 \\
			PMM&0.400 &0.029 &0.400&0.045&0.398 &0.035 &0.393 &0.036 \\
			\hline
	\end{tabular}}
\end{table}

\subsection{More Simulation Studies} \label{sec:moresims}
We investigate the performance of GB-NPMLE under two additional settings for the prior $\pi$: (iv) a trimodal density, and (v) a bounded and skewed left density. Namely, we consider: 
\begin{enumerate}[start=4, label=(\roman*)]
	\item  Gaussian trimodal mixture (\textbf{GMM-tri}): $y \mid \theta \sim \mathcal{N}(\theta, 1)$ and $\theta = 0.2\mathcal{N}(-4,0.5)+0.6\mathcal{N}(0,1)+0.2\mathcal{N}(4,0.5)$;
	\item Binomial-beta mixture (\textbf{BBM}): $y \mid \theta \sim \text{Binomial}(10, \theta)$ and $\theta \sim \text{Beta}(3, 2)$.
\end{enumerate}
\begin{figure}[h]
	\centering
	\includegraphics[width=1.0\columnwidth]{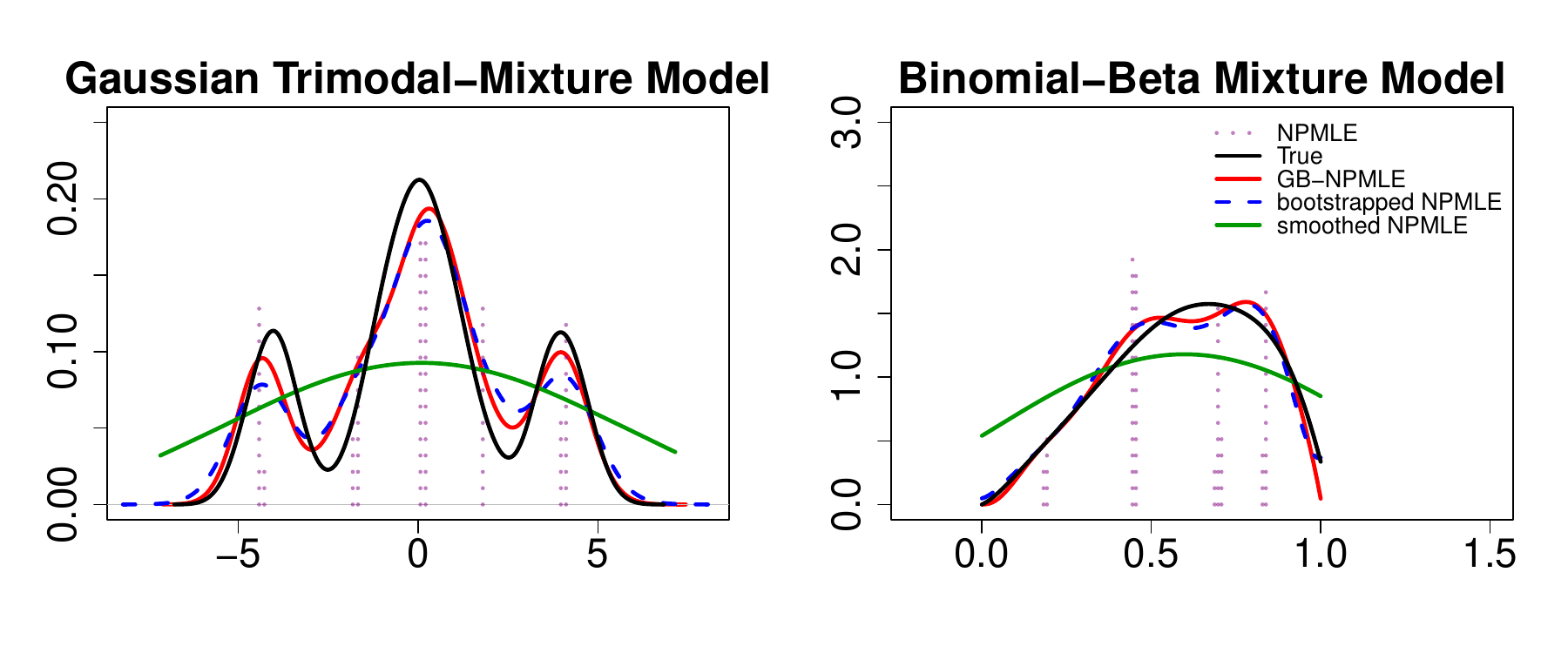}
	\caption{Results from one replication of Simulations (iv)-(v).  In addition to GB-NPMLE (solid red), bootstrapped NPMLE (dashed blue), and smoothed NPMLE (solid green), we also plot the true density (solid black) and the classical discrete NPMLE (dotted purple).}\label{fig:sim_supp}
\end{figure}

As shown in Fig. \ref{fig:sim_supp}, GB-NPMLE approximates the bootstrapped NPMLE quite well for both GMM-tri and BBM. GB-NPMLE and bootstrapped NPMLE also provide density estimates that are close to the true $\pi$, whereas the smoothed NPMLE with bandwidth chosen from LOOCV is once again oversmoothed. A performance comparison of GB-NPMLE, bootstrapped NPMLE, and smoothed NPMLE based on the average of 20 replications is summarized in Table \ref{tab:sim_supp}. 

\begin{table}[tbh]
	\centering
	\caption{COMPARISONS OF PERFORMANCE OF DIFFERENT NPMLE METHODS}\label{tab:sim_supp}
	\resizebox{1.0\columnwidth}{!}{
		\begin{tabular}{c|cc|cc|cc}
			\hline
			&\multicolumn{2}{c|}{GB-NPMLE}&
			\multicolumn{2}{c|}{Bootstrapped NPMLE}&
			\multicolumn{2}{c}{Smoothed NPMLE}\\
			\hline
			Model&$W_1(\pi,\widehat{\pi})$&ISE&$W_1(\pi,\widehat{\pi})$&ISE&$W_1(\pi,\widehat{\pi})$&ISE\\
			\hline
			\hline
			GMM-tri&0.213&0.058&0.273&0.071&0.908&0.031\\
			BBM&0.033&0.059&0.035&0.055&0.109&0.320\\
			\hline
			\hline
	\end{tabular}}
\end{table}

\subsection{Convergence Analysis}\label{sec:cov}
\begin{figure}[h]
	\centering

	\includegraphics[width=\columnwidth]{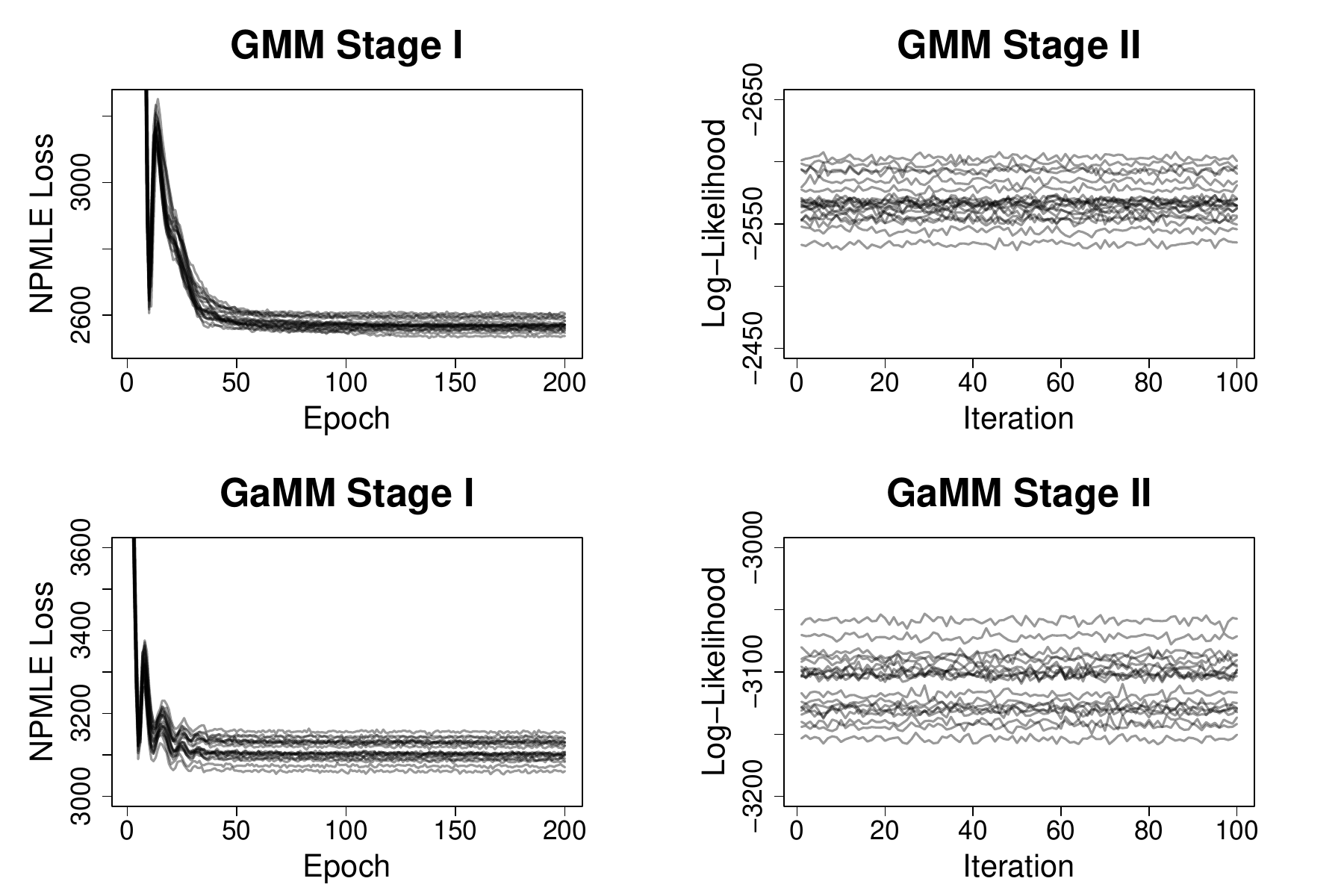}
	\caption{Convergence plots for Stage I and Stage II of the two-stage algorithm under the GMM and GaMM models (Section \ref{sec:experiments}).}\label{fig:cov}
\end{figure}

To empirically check the convergence of the proposed two-stage algorithm, we plot the GB-NPMLE loss vs. epoch number for training the generator $G$ in Stage I. For Stage II, we plot the log-likelihood vs. MCEM iteration number for learning the multinomial weights $\boldsymbol{\tau}$. Fig. \ref{fig:cov} plots these solution paths for all 40 replications of the Gaussian and Gamma mixture models (i.e. Experiments (i) and (ii) in Section \ref{sec:experiments}, denoted as GMM and GaMM respectively. 

Fig. \ref{fig:cov} shows that Stage I  converges quickly, typically within 50 epochs. 
Meanwhile, Stage II converges even faster -- typically within one or two iterations, resulting in a very flat solution path. The convergence plots for our other numerical experiments were very similar to those in Fig. \ref{fig:cov}. Although empirical evidence suggests that the GB-NPMLE two-stage algorithm converges quickly, a rigorous theoretical analysis of convergence rate should be done for future work.

\end{document}